\documentclass[12pt,reqno]{amsart}
\advance\voffset by -2.0cm \advance \hoffset by -1.5cm
\usepackage{hyperref}

\usepackage{cite}
\usepackage{amsmath}
\usepackage{amssymb}
\usepackage{amsthm}
\usepackage{graphicx,color}
\usepackage{verbatim}

\usepackage{mathtools}

\usepackage{bbm}

\newcommand{\CC}{\mathbb{C}} 
\newcommand{\RR}{\mathbb{R}}

\newcommand{\HH}{\mathbb{H}}
\newcommand{\ZZ}{\mathbb{Z}}

\newcommand{\MM}{\mathbb{M}}

\newcommand{\Hh}{\mathcal{H}}

\newcommand{\g}{\mathfrak{g}}
\newcommand{\h}{\mathfrak{h}}

\usepackage{xcolor}

\newtheorem{theorem}{Theorem}

\DeclareMathOperator{\Aut}{Aut}

\DeclareMathOperator{\Tr}{Tr}

\numberwithin{equation}{section}

\def\be{\begin{equation}}
\def\ee{\end{equation}}
\newcommand{\bea}{\begin{eqnarray}}
\newcommand{\eea}{\end{eqnarray}}

\allowdisplaybreaks[3]
\begin{document}
\title[BPS algebras, Genus Zero, and the Heterotic Monster]{BPS algebras, Genus Zero, and the Heterotic Monster}
\author{Natalie M. Paquette}
\address{Stanford  Institute  for  Theoretical  Physics,  Department  of  Physics,  Stanford  University,
Stanford, CA 94305, USA}
\email{npaquett@stanford.edu}

\author{Daniel Persson}
\address{Department of Physics and Department of Mathematical Sciences, Chalmers University of Technology,
412 96, Gothenburg, Sweden}
\email{daniel.persson@chalmers.se}

\author{Roberto Volpato}
\address{Dipartimento di Fisica e Astronomia `Galileo Galilei', Universit\`a di Padova \& INFN, sez. di Padova, Via Marzolo 8, 35131, Padova, Italy}
\email{volpato@pd.infn.it}

\begin{abstract} In this note, we expand on some technical issues raised in \cite{PPV} by the authors, as well as providing a friendly introduction to and summary of our previous work. We construct a set of heterotic string compactifications to 0+1 dimensions intimately related to the Monstrous moonshine module of Frenkel, Lepowsky, and Meurman (and orbifolds thereof). Using this model, we review our physical interpretation of the genus zero property of Monstrous moonshine. Furthermore, we show that the space of (second-quantized) BPS-states forms a module over the Monstrous Lie algebras $\mathfrak{m}_g$---some of the first and most prominent examples of Generalized Kac-Moody algebras---constructed by Borcherds and Carnahan. In particular, we clarify the structure of the module present in the second-quantized string theory. We also sketch a  proof of our methods in the language of vertex operator algebras, for the interested mathematician.
\end{abstract}
\maketitle

\tableofcontents

\section{Introduction}

The Monstrous moonshine conjecture, initiated by a series of observations by John McKay and Thompson in the 70s and precisely formulated by Conway and Norton \cite{Conway:1979kx}, is the arena of some of the most fruitful cross-breeding between physics and mathematics.  It motivated the development of several brand new areas in algebra, number theory, and group theory, such as vertex operator algebras and Generalized (Borcherds) Kac-Moody algebras. The intuition for many of these developments came from physics, in particular two dimensional conformal field theory (CFT) and string theory. The conjecture has been proven by Borcherds \cite{Borcherds}, with some fundamental contributions by Frenkel, Lepowsky and Meurman (FLM) \cite{FLM}, among others. While some of the tools used in the proof were explicitly inspired by CFT and string theory, the physical interpretation of many aspects of Monstrous moonshine remains mysterious. Our work \cite{PPV} is an attempt to fill in this gap.

The moonshine conjecture can be formulated as follows. There exists a graded vector space $V^\natural=\oplus_{n=-1}^\infty V^\natural_{n}$, such that 
\begin{enumerate}
\item each $V^\natural_{n}$ is a finite dimensional representation of the Monster group $\MM$
\item for each $g\in \MM$, the $q$-series (McKay-Thompson series)
\be T_{1,g}(\tau)=\sum_{n=1}^\infty \Tr_{V^\natural_n} (g) q^n\ ,\qquad q=e^{2\pi i\tau}\ ,
\ee converges on the upper half-plane $\HH$ to a modular function under some discrete subgroup $\Gamma_g\subset SL(2,\RR)$, i.e.
\be T_{1,g}(\frac{a\tau+b}{c\tau+d})=T_{1,g}(\tau)\ ,\qquad \begin{pmatrix}
a & b\\ c & d
\end{pmatrix}\in \Gamma_g\ ,
\ee
\item the group $\Gamma_g$ is genus zero (i.e., the compactified quotient $\hat\HH/\Gamma_g$ has the topology of a sphere) and $T_{1,g}$ is a Hauptmodul, i.e. it establishes a biholomorphic map from $\hat\HH/\Gamma_g$ to the Riemann sphere $\CC\cup \{\infty\}$.
\end{enumerate}
In \cite{FLM}, Frenkel, Lepowsky, and Meurman showed that $V^\natural$ can be interpreted as the space of states of a chiral CFT (or holomorphic VOA) with central charge $24$, whose automorphism group is the Monster group $\MM$. The McKay-Thompson series are therefore interpreted as `twined' partition functions
\be T_{1,g}(\tau)=\Tr_{V^\natural}(gq^{L_0-1})\ ,
\ee where $L_0$ is the Virasoro generator. In physics, $T_{1,g}$ is given by a path integral over a torus with modular parameter $\tau$ with the fields obeying $g$-twisted periodicity conditions along one of the cycle of the torus. From a physicist's viewpoint, it is then natural to expect $T_{1,g}$ to be invariant   under the subgroup of $SL(2,\ZZ)$ that respects these periodicity conditions, due to the invariance of the theory under some large conformal transformations on the torus. In fact, one can also consider more general twisted-twining partition functions
\be T_{g,h}(\tau)=\Tr_{V^\natural_g}(hq^{L_0-1})\ee for each commuting pair $g,h\in \MM$, given by twisted periodicity conditions on two independent cycles on the torus. Here, the trace is taken over the $g$-twisted sector $V^\natural_g$ of the CFT. A generalized moonshine based on $T_{g,h}$ was proposed by Norton \cite{Norton1987}.

What singles out the Monstrous moonshine from analogous CFT constructions is the Hauptmodul property. A crucial point for this property to hold is that the groups $\Gamma_g\subset SL(2,\RR)$ contain, in general, elements that are not in $SL(2,\ZZ)$ and are therefore larger than the ones expected by purely CFT arguments -- in general, CFT groups are not genus zero. The genus zero property was finally proved by Borcherds \cite{Borcherds}, using the existence of a Generalized, or Borcherds, Kac-Moody algebra (GKM) with automorphism group $\MM$.  The construction of the algebra is based on the FLM conformal field theory and is directly inspired by a bosonic string construction. However, a natural physical interpretation of the Hauptmodul property was not clear. In particular,  there is no physical explanation of the elements of $\Gamma_g$ not contained in $SL(2,\ZZ)$ (Atkin-Lehner involutions).

In our work \cite{PPV}, we propose a possible physical interpretation of the Monstrous moonshine conjecture in the context of heterotic strings. For each $g\in \MM$, we consider a certain heterotic string compactification to $0+1$ dimensions. Each such model has a GKM algebra of `spontaneously broken gauge symmetries', generated by certain BRST-exact states in the string theory. When $g$ is the identity, this Kac-Moody algebra is exactly Borcherds' Monster Lie algebra. Then, we define supersymmetric indices $Z_{g,1}(T,U)$, depending on some moduli $(T,U)\in\HH\times \HH$ of our compactification, and counting supersymmetric (BPS) states in these heterotic models. These indices are directly related to the McKay-Thompson series via the identity
\be Z_{g,1}(T,U)=(T_{1,g}(T)-T_{g,1}(U))^{24}\ .
\ee In this context, modularity of the McKay-Thompson series under the groups $\Gamma_g$ described by Conway and Norton is simply a consequence of the invariance of $Z_{g,1}(T,U)$ under the T-dualities of the model. Furthermore, the Hauptmodul property of $T_{1,g}$ follows from a careful analysis of the behaviour of $Z_{g,1}$ at the boundary of the moduli space.

In this article, after a quick review of \cite{PPV}, we provide a detailed explanation of two key steps of this construction. First of all, we describe the explicit realization of the action of the GKM algebra on the BPS states contributing to the index $Z_{g,1}$. Furthermore, we provide an interpretation of the Atkin-Lehner dualities from the point of view of vertex operator algebras. The latter is one of the key point in our interpretation of the Hauptmodul property of Monstrous moonshine.

\section{Our models}
\label{sec:models}

\noindent In this section, we outline the heterotic compactifications that will be our focus in the sequel. Details of the construction are provided in \cite{PPV}. Begin by considering a compactification of the heterotic string to 1+1 dimensions. We choose the internal CFT of central charges $(c, \bar{c}) = (24, 12)$ to be the product $V^{\natural} \times \bar{V}^{s\natural}$, where the former is the Frenkel, Lepowsky, Meurman (FLM) Monster module \cite{FLM} and the latter is the Conway moonshine module described by FLM and studied in detail by Duncan \cite{duncan2007super}. The Monster module will be particularly relevant to our discussion, and we may think of it as a compactification of 24 chiral bosons on the Leech lattice, followed by an orbifold by the canonical $\mathbb{Z}_2$ symmetry that negates all lattice vectors; see \cite{DGH} for a physics discussion of this model. The partition function for the FLM theory is the famous modular J-function with zero constant term:
\begin{equation}
Tr_{V^{\natural}}(q^{L_0 - 1}) = J(\tau) = q^{-1} + 0 + 196884 q + \ldots, \qquad q:=e^{2\pi i \tau}.
\end{equation} 
The Conway module largely plays the role of a spectator in our story, but we make use of the following properties: it has $\mathcal{N}=1$ supersymmetry, its Ramond sector has 24 ground states of conformal weight $1/2$ with positive fermion number, and its NS sector has no fields of conformal weight $1/2$. Intuitively, one may think of this relatively exotic-looking construction as an asymmetric $\mathbb{Z}_2 \times \mathbb{Z}_2$\footnote{The $\mathbb{Z}_2$ orbifold that is part of the construction of the Conway module flips the signs of the 8 right-moving scalar superfields.} orbifold of a $\mathbb{T}^8$ compactification, where we have tuned the moduli of the latter to the unique point such that the internal CFT factorizes in the desired way. It turns out that this model has $(0, 24)$ space-time supersymmetry \cite{BergmanDistler}.

More generally, we want to consider a set of orbifolds of this model, using an orbifolding procedure analogous to that used in the construction of CHL models \cite{Chaudhuri:1995fk, Chaudhuri:1995ee, Chaudhuri:1995bf, Chaudhuri:1995dj}. More precisely, we compactify the above model on an additional $S^1$ of radius $R$. We then orbifold the model by a $\mathbb{Z}_N$ symmetry $(\delta, g)$, where $\delta$ is a shift by $1/N$-th of a period along the $S^1$ and $g$ is an order $N$ symmetry of the Monster group, which acts on the FLM module. It turns out that the resulting CHL-like model only depends on the conjugacy class of $\langle g \rangle$. If the element $g$ does not satisfy the level-matching condition we can still construct consistent orbifolds of order $N\lambda$ where $\lambda >1$ measures the failure of level-matching; see \cite{Persson:2015jka, PPV} for details of this slightly more subtle case. In the remainder of this note we will focus on the $\lambda=1$ case for ease of exposition.

We mention in passing that these models have subtleties related to gravitational anomalies that are discussed in \cite{PPV}; to circumvent these difficulties, we will always work at zero string coupling, $g_s=0$.

\section{Monstrous algebras}
\label{sec:algebras}
The Monster Lie algebra $\mathfrak{m}$  was introduced by Borcherds in his proof of the Monstrous moonshine conjectures \cite{Borcherds}, and its `twisted' counterparts $\mathfrak{m}_g$ by Carnahan in his proof of generalized Monstrous moonshine \cite{2008arXiv0812.3440C,Carnahan2014,2012arXiv1208.6254C}. Given the presence of certain string-theoretic ingredients in Borcherds' proof (e.g. the addition of lightcone directions and the use of the no-ghost theorem), it is natural to conjecture that these Generalized Kac-Moody algebras (GKMs) \footnote{GKMs or Borcherds-Kac-Moody algebras (BKMs) are generalizations of (super)Kac-Moody algebras that allow for imaginary simple roots. The ones that appear in Monstrous moonshine are infinite dimensional examples of rank two.} would play a role in a full string theory uplift of the Monster module, as has been suggested in e.g. \cite{Carnahan2014}\footnote{GKMs have also been proposed, by Harvey and Moore, to be the structure relevant for studying algebras of BPS states in 4d $\mathcal{N}=2$ compactifications\cite{Harvey:1995fq,Harvey:1996gc}. See \cite{Neumann:1997pr,Fiol:2000wx,Kontsevich:2010px,Gukov:2011ry,Chuang:2013wt} for other interesting work elucidating the structure of algebras of BPS states.}. We argue that the spacetime BPS states in our theories furnish a module over (or representation of) the associated GKM $\mathfrak{m}_g$, similarly to the discussions in \cite{Gaberdiel:2011qu,Hohenegger:2011ff}, and that $\mathfrak{m}_g$ plays the role of a spontaneously broken gauge algebra. 

To understand this, let us start by considering ordinary first-quantized string theory. In particular, we want to understand the single string BPS states in our theories. We will focus on the unorbifolded theory, but the basic analysis will carry through for our orbifold models as well. Essentially, the superpartners of the BPS states turn out to be $\mathcal{Q}$-exact with respect to the right-moving part of the nilpotent BRST operator. Moreover, massless BRST-exact states in string theory generate gauge transformations. In the context of our low-dimensional (0+1) compactification, this interpretation has subtleties, coming from the fact that we can no longer identify the mass of states with the momentum-squared, $k^2$. However, since our BRST-exact states will always carry a non-zero momentum,  we anticipate that we may roughly think of them as (behaving like) massive null states. These states will still generate an algebra that turns out to be isomorphic to $\mathfrak{m}_g$ and so we interpret $\mathfrak{m}_g$ heuristically as a spontaneously broken gauge symmetry. It would of course be interesting to understand if and when the gauge symmetry would be restored in our models, as in some appropriate tensionless limit of the string theory. 

Let us now be slightly more explicit, referring as always to \cite{PPV} for the many details we will gloss over. The BPS states are physical states in the Ramond sector with right-moving momentum and conformal weight $k_R^2 = 0, h_R = 1/2$. Recall that, essentially due to the properties of $\bar{V}^{s\natural}$ listed above, we will have 24 copies of any given BPS state and that our BPS states are fermionic. One can label the BPS state as $| \chi, i, \alpha, k \rangle$, with $\chi \in V^{\natural}$ of conformal weight $h_{\chi}$, $i= 1, \ldots, 24$ denotes one of the 24 Ramond ground states of $\bar{V}^{s\natural}$, $\alpha$ denotes a Ramond ground state of positive chirality in the spacetime direction which is tensored with the state labeled by $i$, and the space-time momentum $k$ which satisfies level-matching. Computing the action of spacetime supersymmetry on our BPS state, we find that its superpartner $V(z, \bar{z})$(written explicitly in \cite{PPV} as an unintegrated vertex operator in the $-1$-picture) satisfies \begin{equation}
V(z, \bar{z}) = \left\lbrace \mathcal{Q}^{BRST}_R, W(z, \bar{z}) \right\rbrace, \ \left\lbrace \mathcal{Q}^{BRST}_L, W(z, \bar{z}) \right\rbrace = 0
\end{equation} for a certain operator $W(z, \bar{z})$ given in \cite{PPV}; $L, R$ denote left and right-moving parts of the full BRST charge. We next denote the integrated $0$-picture form of $V(z, \bar{z})$ as $\mathcal{W}_{\chi}$ and examine the consequences of inserting it into a string amplitude $\langle \mathcal{W}_{\chi} \mathcal{V}_1 \mathcal{V}_2 \ldots \rangle$, where the $\mathcal{V}_i$ are vertex operators associated with BPS states. We compute the action of this insertion on the vertex operators in the correlation function and find that it is given by the zero-modes of holomorphic currents $\mathcal{V}_{\chi} e^{i k_L X_L}$. These zero-modes generate precisely the Monster Lie algebra. We also see from the action of $\mathcal{W}_{\chi}$ on the single particle BPS states that the latter indeed form a module over this algebra. 

It is instructive to note that in Borcherds' proof of the Monstrous moonshine conjectures, $\mathfrak{m}$ comes from studying the cohomology of a certain BRST-like operator acting on the vertex operator algebra given by $V^{\natural} \times \Gamma^{1, 1}$, with $\Gamma^{1, 1}$ the unimodular lattice of signature $(1, 1)$. In our model, Borcherds' operator is simply $\mathcal{Q}^{BRST}_L$, and imposing level-matching and right-moving mass-shell conditions forces the left-moving momentum $k_L$ to take values in a lattice isomorphic to $\Gamma^{1, 1}$. The left-moving mass-shell condition is then simply Borcherds' physical state condition.

We next want to examine the second-quantized version of our string model, where we allow an arbitrary number of fundamental strings and study the resulting Fock space. In order to ultimately  understand the genus 0 properties of Monstrous moonshine, we will compute and study the supersymmetric indices of these second-quantized theories extensively in the sequel. For now, let us examine the algebraic properties enjoyed by the second-quantized spacetime BPS states. We have shown that each Monstrous orbifold model contains an infinite dimensional Lie algebra given by $\mathfrak{m}_g$ and each generator of the algebra has an associated 24 fermionic spacetime BPS states that form 24 copies of the adjoint representation. We now consider the free theory limit of the second quantized BPS Hilbert space $\mathcal{H}_{BPS}$ as a Fock space built from fermionic oscillators (corresponding to our spacetime BPS states) acting on the vacuum \footnote{For now, we avoid the issue of vacuum degeneracy that occurs at the self-dual radius of $S^1$.} in all possible ways. If we focus on just one of the 24 copies, we may associate a fermionic mode $\eta_a$ to each $a \in \mathfrak{m}_g$ and derive the anticommutation relations
\begin{equation}
\left\lbrace \eta_a, \eta_b \right\rbrace = 0, \ a,b \in \mathfrak{m}_g.
\end{equation} If we consider the usual triangular decomposition of our algebra $\mathfrak{m}_g = \mathfrak{g}^- \oplus \mathfrak{h} \oplus \mathfrak{g}^+$ we see, by defining the action of the oscillators on the ground state as $\eta_a |0 \rangle = 0, \ a \in \mathfrak{g}^+ \oplus \mathfrak{h}$, that the Fock space is isomorphic to $\bigwedge \mathfrak{g}^-$ and inherits the natural weight space grading from $\mathfrak{m}_g$: $\mathcal{H}_{BPS} = \bigoplus_{\gamma \in \Gamma}\mathcal{H}_{\gamma}$.

\section{The action of the Monstrous algebras on BPS states}

The space $\Hh_{BPS}$ of BPS states contributing to the supersymmetric index is naturally isomorphic to $\bigwedge \g^-$. If instead of $\mathfrak{m}_g$ we were considering an ordinary Kac-Moody algebra, it would immediately follow that $\bigwedge \mathfrak{g}^-$ is an irreducible highest weight module for the algebra with the highest weight given by the Weyl vector \cite{kostant}. For GKMs, it is unknown whether $\bigwedge \mathfrak{g}^-$ is a (possibly reducible) module. However, for the particular GKM algebras $\mathfrak{m}_g$ that are relevant for our construction, we are able to determine explicitly the action of the algebra on $\bigwedge \g^-$.

The construction is as follows. The algebra $\g$ has a two-dimensional Cartan subalgebra $\h$ and is graded by the natural weight lattice $\Gamma\subset \h^*$, $\g=\oplus_{\gamma\in\Gamma} \g^\gamma$. It is endowed with an invariant non-degenerate bilinear form $\kappa$, such that, for any $a\in \g^\gamma$, $\kappa(a,b)$ vanishes unless $b\in \g^{-\gamma}$. 
Since $\kappa$ is invariant under the adjoint action of $\g$ on itself, one can identify $\g$ with a subalgebra of the orthogonal algebra $\mathfrak{o}(\kappa)$, the Lie algebra associated with the orthogonal group preserving $\kappa$. Roughly speaking, $\bigwedge\g^-$ can be interpreted as a spinor representation of this orthogonal algebra $\mathfrak{o}(\kappa)$ and therefore of its subalgebra $\g$.\footnote{We thank Howard Garland for suggesting this point of view to us.}  In the rest of this section, we will make this idea precise for the infinite dimensional algebras we are interested in.

We start by defining the Fock space $\Hh_{BPS}$ in a slightly different fashion than in \cite{PPV}. We associate with each element $a\in \g$ a fermionic operator $\xi_a$, satisfying the anti-commutation relations
\be \{\xi_a,\xi_b\}=\kappa_{ab}\ .
\ee This is simply the Clifford algebra associated with the bilinear form $\kappa_{ab}$. In particular, since the Cartan subalgebra $\h$ is a two-dimensional real space with signature $(1,1)$, we can take a light-cone decomposition $\h=\h^+\oplus\h^-$ into two one-dimensional null subspaces $\h^+,\h^-$, such that the corresponding algebra operators $\xi_+,\xi_-$ satisfy
\be \{\xi_+,\xi_-\}=1\qquad \{\xi_+,\xi_+\}=0=\{\xi_-,\xi_-\}\ .
\ee We stress that the operators $\xi_a$ are different from the operators $\eta_a$ defined in \cite{PPV}, where all anti-commutators were set to zero. The precise relation is $\eta_a=\xi_a(\xi_--\xi_+)$ for $a\in \g^-$ and $\eta_a=0$ for $a\in \h\oplus \g^+$. 
Next, let us define a vacuum state $|0\rangle$ such that
\be \xi_a|0\rangle=0\qquad \forall a\in \g^+\oplus \h^+\ ,
\ee and define a Fock space $S$ by acting on $|0\rangle$ in all possible ways with the operators $\xi_a$, $a\in \g^-\oplus \h^-$.  Finally, we introduce a `chirality' operator $(-1)^J$ satisfying
\be (-1)^J|0\rangle=|0\rangle\qquad [(-1)^J,\xi_a]=-\xi_a\qquad \forall a\in \g\ ,
\ee and consider the eigenspace $S_+$ of $(-1)^J$  with positive eigenvalue, i.e. $(-1)^J_{\rvert S_+}=+1$. It is easy to see that the space $S_+$ is isomorphic to $\bigwedge \g^-$, and that the isomorphism
\be \bigwedge^{2n} \g^-\ni a_1\wedge \ldots \wedge a_{2n}\mapsto \xi_{a_1}\cdots \xi_{a_{2n}}|0\rangle \in S_+
\ee
\be \bigwedge^{2n+1} \g^-\ni a_1\wedge \ldots \wedge a_{2n+1} \mapsto \xi_{a_1}\cdots \xi_{a_{2n+1}}\xi_-|0\rangle\in S_+\ ,
\ee preserves the grading $S_+=\bigoplus_{\gamma\in \Gamma}S_+^\gamma$ by the weight lattice $\Gamma$. The operator $(-1)^J$ should not be confused with the operator $(-1)^F:=1-2\xi_-\xi_+$ (that was called the fermion number in \cite{PPV}), acting on $S_+\cong\bigwedge \g^-$ by
\be (-1)^Fa_1\wedge\ldots \wedge a_n= (-1)^na_1\wedge\ldots \wedge a_n\ .
\ee The fermion number $(-1)^F$ will play a role in the definition of the second quantized index in section \ref{sec:index}.

Let us show that the space $S_+$ affords us a representation of the algebra $\g$. Let $\Sigma:\g\to \mathfrak{gl}(S_+)$  be the linear map defined by
\be\label{generator1} \Sigma(a):=\frac{1}{2}\sum_{b,c\in \g} \kappa^{bc}\xi_{[a,b]}\xi_c \ ,\qquad a\in \g^+\oplus\g^-
\ee 
\be\label{generator2} \Sigma(\mu):=\sum_{\substack{c\in \g^+\oplus \h^+\\ b\in \g^-\oplus \h^-}} \kappa^{bc}\xi_{[\mu,b]}\xi_c -\rho(\mu)\ ,\qquad \mu\in \h
\ee
where  $\kappa^{bc}$ denotes the inverse of the metric $\kappa$ and $\rho\in \h^*$ is a suitable weight. 
Notice that although the expressions \eqref{generator1} and \eqref{generator2}  involve infinite sums, only finitely many terms are non-vanishing when acting on a state $|\psi\rangle\in S_+$ of definite weight $\gamma$. A direct calculation shows that
\be\label{adjaction} [\Sigma(a),\xi_b]= \xi_{[a,b]}\ ,
\ee  
for all $a,b\in \g$. Since the only operators commuting with all $\xi_a$ are the constants, \eqref{adjaction} determines the operators $\Sigma(b)$ for all $b\in \g$ up to constants. Eq.\eqref{adjaction} also implies that
\be\label{commutation} [\Sigma(a),\Sigma(b)]- \Sigma([a,b])=\ell(a,b)\ ,
\ee for some constants $\ell(a,b)$, because the left-hand side commutes with all $\xi_c$, $c\in \g$. In order to show that $\Sigma$ is a representation of $\g$, we need to show that, for a suitable choice of $\rho\in \h^*$, the constants $\ell(a,b)$ vanish for all $a,b\in \g$. By \eqref{commutation},  the constants $\ell$ satisfy
\be\label{ellrel}  \ell([a,b],c)+\ell([b,c],a)+\ell([c,a],b)=0\ ,
\ee for all $a,b,c\in \g$. Furthermore, if $a\in \g^\gamma$, $\ell(a,b)$ vanishes unless $b\in \g^{-\gamma}$, because only in this case the operator on the left-hand side of \eqref{adjaction} preserves the weight of a state. Finally, if $a,b\in \h$, then $\ell(a,b)=0$, as is clear by acting on $|0\rangle$ with the operator on the left-hand side of \eqref{commutation}. By \eqref{ellrel}, if some $x,y\in \g^-$ satisfies $\ell(x,a)=0=\ell(y,a)$ for all $a\in \g^+$, then $\ell([x,y],a)=0$ for all $a\in \g$. Hence, a sufficient condition for $\ell(a,b)$ to vanish for all $a,b\in \g$ is that $\ell(a,b)=0$ for all $a\in \g^{\gamma}$, $b\in \g^{-\gamma}$ and for all \emph{simple} roots $\gamma$. 

Let $\gamma\in \Gamma$ be a simple root and consider $b\in \g^{-\gamma}$ and $a\in \g^\gamma$.  Since $\gamma$ is simple,
 we have
\be \Sigma(b)|0\rangle=\sum_{\mu,\nu\in \h}\kappa^{\mu\nu}\gamma(\mu) \xi_b\xi_\nu|0\rangle\qquad \Sigma(a)|0\rangle=0\ .
\ee By \eqref{adjaction}, it follows immediately that
\be [\Sigma(a),\Sigma(b)]|0\rangle= \frac{1}{2}\kappa(\gamma,\gamma) \kappa(a,b)\ .
\ee 
On the other hand, we have
\be \Sigma([a,b])|0\rangle= -\kappa(a,b)\kappa(\rho,\gamma)\ .
\ee Therefore, $\ell$ vanishes identically if and only if the vector $\rho$ satisfies
\be \kappa(\rho,\gamma)=-\frac{1}{2}\kappa(\gamma,\gamma)\ ,
\ee for all simple roots $\gamma$. This is nothing but the defining property of the Weyl vector. Therefore, assuming that the algebra $\g$ admits a Weyl vector, we proved that the space $S_+\cong \bigwedge \g^-$, which is isomorphic to the space of BPS states, carry a representation $\Sigma$ of the algebra $\g$, with lowest weight the Weyl vector $\rho$. We stress that, in this proof, we used repeatedly the fact that the Cartan algebra $\h$ is two dimensional. We also assumed that a Weyl vector for the algebra $\g$ exists. Both these properties are satisfied for the Monstrous GKM algebras $\mathfrak{m}_g$.

\section{The second-quantized index}
\label{sec:index}

With the Fock space construction of the previous section in hand, it is natural to compute a supersymmetric index that provides a signed count of spacetime BPS states in the second quantized models. The results gleaned from these computations will prove crucial for our physics interpretation of the genus zero property in the following section. We can readily define the index in the unorbifolded model from the Fock space construction as
\begin{equation}
Z(T, U):= Tr_{\mathcal{H}_{BPS}}\left((-1)^F e^{2 \pi i T W} e^{2 \pi i U M} \right)
\end{equation} 
where $F$ is the fermion number, $(W, M)$ are winding and momentum operators along the $S^1$, and $(T, U) \in \mathbb{H} \times \mathbb{H}$ are their associated complexified chemical potentials, which are functions of the radius $R$ and the inverse temperature $\beta$. More precisely, they are
\begin{equation}
T = b + i \frac{\beta R}{2 \sqrt{2} \pi}, \ U = v + i \frac{\beta}{2 \sqrt{2} \pi R}
\end{equation} in terms of the real chemical potentials $b, v$ conjugate to winding and momentum, respectively. 
For the orbifolded models, one can analogously define 
\begin{equation}
Z_{g, 1}(T, U):= Tr_{\mathcal{H}^g_{BPS}}\left((-1)^F e^{2 \pi i T W} e^{2 \pi i U M} \right)
\end{equation}
where $\mathcal{H}^g_{BPS}$ is the space of BPS states of the orbifolded model (i.e. $\bigwedge \mathfrak{m}_g^{-}$, or $\bigwedge \g^{-}$ in the notation of the previous section). Even more generally, for elements of the Monster $h$ that commute with $g$ one may define `twisted-twined' indices
\begin{equation}
Z_{g, h}(T, U):= Tr_{\mathcal{H}^g_{BPS}}\left(h(-1)^F e^{2 \pi i T W} e^{2 \pi i U M} \right).
\end{equation}
One can motivate these expressions by starting with a trace over the full Hilbert space and using appropriate index arguments to show that the trace will localize to the subsector spanned by short multiplets. From the Fock space description it is straightforward to compute the answer first in the unorbifolded model,
\begin{equation}
Z(T, U) = e^{24 (2\pi i (m_0 U + w_0 T))} \prod_{\substack{m> 0\\w\in \ZZ}} \left(1 - e^{2 \pi i U m} e^{2 \pi i T w} \right)^{24 c(m w)},
\end{equation} where we have allowed for vacuum winding and momentum $(w_0, m_0)$ that need to be fixed by other means. Here, $c(n)$ are the coefficients of $J(\tau) = \sum_{n=-1}^{\infty} c(n) q^n$. 

In the orbifolded case (considering as always $g$ of order $N$ and with $\lambda=1$), one instead computes
\begin{equation}
Z_{g, 1}(T, U) = e^{24 (2\pi i (m_0 U + w_0 T))} \prod_{\substack{ n> 0\\ m \in \ZZ}} \left(1 - e^{2 \pi i U m/N} e^{2 \pi i T n} \right)^{24 \hat{c}(m n/N)}
\end{equation} 
where now $\hat{c}(mn/N)$ are the dimensions of the $e^{2 \pi i m/N}$-eigenspace of the $g^n$-twisted sector $V^{\natural}_{g^n}$ at level $L_0 - 1 = mn/N$ \footnote{More formally, they are coefficients of the discrete Fourier transforms of the generalized McKay-Thompson series: $F_{l, k}(z) = \frac{1}{N}\sum_{j \in \mathbb{Z}/N\mathbb{Z}} e^{- 2 \pi i \frac{jk}{N}} T_{g^l, g^j}(z) = : \sum_{n \in \mathbb{Z}/N\mathbb{Z}} \hat{c}_{l, k}(n) e^{2 \pi i z n}$.} and we have defined
\begin{equation}
T = b + i \frac{\beta R}{2 \sqrt{2} \pi N}, \ U = v + i \frac{\beta N}{2 \sqrt{2} \pi R}.
\end{equation}
It will be instructive for us to compute the supersymmetric index by a different method, in particular as an appropriate one-loop integral which, among other things, makes the T-duality of the problem manifest. As before, we consider compactified Euclidean time with period $\beta$. This formulation endows $T, U$ with an interpretation as the K{\"a}hler and complex moduli of the torus, respectively. We expect the index to take the form
\begin{equation}\label{oneloop1}
Z_{g, 1}(T, U) = e^{-(S_{tree} + S_{1-loop} + \ldots)}
\end{equation}  
and we moreover expect the index to be 1-loop exact since we are studying a free theory $g_s=0$. The one-loop contribution to the path integral is given by the usual integral over the modular fundamental domain $\mathcal{F}$
\begin{equation}\label{oneloop2}
S_{1-loop} = \frac{1}{2}\int_{\mathcal{F}}\frac{d^2\tau}{\tau_2} \left(Tr_{NS}(q^{L_0-c/24} \bar{q}^{\bar{L_0}- \bar{c}/24}P_{GSO}) - Tr_{R}(q^{L_0-c/24} \bar{q}^{\bar{L_0}- \bar{c}/24}P_{GSO}) \right).
\end{equation}
Though there turn out to be subtleties implementing the GSO projection \cite{PPV}, one can deduce that one recovers $Z(T, U)$ and its (twisted, twined) counterparts in this way. After evaluating the one-loop integral explicitly using standard techniques, one finds
\begin{equation}
Z_{g, 1}(T, U) = e^{24 (-2\pi i T)} \prod_{\substack{ n> 0\\ m \in \ZZ}} \left(1 - e^{2 \pi i U m/N} e^{2 \pi i T n} \right)^{24 \hat{c}(m n/N)}\ .
\end{equation} Notice that this result fixes the vacuum winding/momentum to be $(m_0, w_0) = (0, -1)$. 

As a final point, we notice that (the 24th roots of) these formulae are precisely the denominators for the Monstrous Lie algebras $\mathfrak{m}_g$ and one can compute the index directly in the algebraic formulation. In particular, in the language of section \ref{sec:algebras}, $W$ and $M$ correspond to the generators $\Sigma(\mu)$ along two null directions $\mu\in \h$ spanning the Cartan subalgebra $\h$, and $(w_0,m_0)$ are the corresponding components of the Weyl vector $\rho$. This method enables us to reproduce for these algebras the twisted and twined analogues of the famous Koike-Norton-Zagier formula $Z(T, U) = J(T) - J(U)$, namely
\begin{equation}\label{ZtoMcKay}
Z_{1, g}(T, U)^{1/24} = T_{1, g}(T) - T_{1, g}(U)
\end{equation} and
\begin{equation}\label{ZtoMcKay2}
Z_{g, 1}(T, U)^{1/24} = Z_{g, 1}(T, -\frac{1}{U})^{1/24} = T_{1, g}(T) - T_{g, 1}(U).
\end{equation} We refer the reader to \cite{PPV} for details. 

\section{A physics view on genus 0}
\label{sec:genus0}

Monstrous moonshine is the conjecture (now theorem, by Borcherds) that the McKay-Thompson series $T_{1,g}$ are Hauptmoduln for some genus zero group $\Gamma_g$. The relations \eqref{ZtoMcKay}-\eqref{ZtoMcKay2} between the second quantized supersymmetric index and the McKay-Thompson series allow us to reinterpret the Monstrous moonshine in terms of properties of $Z_{g,1}$.

First of all, modularity of $T_{1,g}$ is directly related to the invariance of $Z_{g,1}$ under T-dualities. The automorphic properties of $Z_{g,1}$ are manifest in its description \eqref{oneloop1},\eqref{oneloop2} in terms of the exponential of a 1-loop integral of the string model at finite temperature $\beta$. One may view this as a compactification of the heterotic on $\mathbb{T}^2 \times V^{\natural} \times \bar{V}^{s\natural}$ or orbifolds thereof, where the Euclidean time has period the inverse temperature $\beta$. For CHL models, the groups of T-dualities can be found in \cite{PPV} and, more generally, in \cite{nuovo}, following the techniques developed in \cite{Persson:2015jka}. Roughly speaking, we study automorphisms of the lattice $L$ of winding-momenta associated to the (orbifolded) $\mathbb{T}^2$, which will be a certain discrete subgroup of $O(2, 2; \mathbb{R})$. For the case $\lambda=1$, the lattice $L$ of a Monstrous orbifold associated to a symmetry $g$ of order $N$ is spanned by
\begin{equation}
(m_1, w_1, m_2, w_2) \in \mathbb{Z}\oplus \frac{1}{N} \mathbb{Z} \oplus \mathbb{Z} \oplus \mathbb{Z}
\end{equation}
and we find that the group of automorphisms $SO^+(L)$ is generated by adjoining the so-called Atkin-Lehner involutions $(W_e, W_e), e \vert \vert N$ (i.e. $e$ is an exact divisor of $N$) to the group $\Gamma_0(N) \times \Gamma_0(N)$, where 
\be \Gamma_0(N):=\left\lbrace\begin{pmatrix}
a & b\\ c & d
\end{pmatrix}\in SL(2,\ZZ)\mid c\equiv 0\mod N\right\rbrace ,
\ee
and 
\be W_e = \frac{1}{\sqrt{e}}\begin{pmatrix}
ae & b \\ cN & de
\end{pmatrix}\in SL(2,\RR),\qquad a,b,c,d,e\in \ZZ.\ee
Similar but more intricate results hold for the case $\lambda>1$. In general, the group $SO^+(L)$ of T-dualities will relate a CHL model at one point in moduli space to a possibly different model with different values of the moduli. The subgroup $G_g\subseteq SO^+(L)$ of the T-duality group that consists of self-dualities is particularly interesting for us. It is generated by $\Gamma_0(N)\times \Gamma_0(N)$ together with a \emph{subset} of all possible Atkin-Lehner involutions. In particular, one must include  $(W_e,W_e)$ for the exact divisors $e$ such that the orbifold of $V^\natural$ by $g^e$ has no currents. It is proved in \cite{PPV} that in this case the orbifold $V^\natural/\langle g^e\rangle$ is necessarily isomorphic to $V^\natural$.  The index $Z_{g,1}(T,U)$ must be invariant (possibly up to some phase) under the action of the group $G_g$ of self-dualities. Given the relation \eqref{ZtoMcKay}, this implies that the McKay-Thompson series $T_{1,g}(T)$ are invariant under the projection ${\rm proj_1}(G_g)$ of $G_g\subset SL(2,\RR)\times SL(2,\RR)$ onto the the first $SL(2,\RR)$ factor. It turns out that these self-duality groups are exactly the groups $\Gamma_g$ predicted by the Monstrous moonshine conjecture. Modularity of $T_{1,g}$ under the group $\Gamma_0(N)\subseteq SL(2,\ZZ)$ can also be derived by its description in terms of a `twined' partition function in 2D CFT. However, there is no way to understand the action of the Atkin-Lehner involutions in this picture. Therefore, our description provides the first complete physical interpretation of the modularity properties of the McKay-Thompson series.

The Atkin-Lehner involutions play a crucial role in one the most surprising properties of Monstrous moonshine, namely the observation that the invariant subgroup $\Gamma_g$ is always genus zero (i.e., the compactified quotient $\hat\HH/\Gamma_g$ is a sphere) and the McKay-Thompson series $T_{1,g}$ is  the generator of the field of meromorphic functions on this sphere (a Hauptmodul). Our construction provides a clean physical interpretation of this property. Let us sketch the heuristic idea behind this interpretation; we refer to \cite{PPV} for the details. First of all, the Hauptmodul property can be reformulated as the claim the $T_{1,g}(\tau)$ has only a single pole at $\tau\to i\infty$ and at its images under $\Gamma_g$. This automatically implies that $\Gamma_g$ is a genus zero group with Hauptmodul $T_{1,g}$, since $T_{1,g}$ defines a one-to-one holomorphic map from $\hat \HH/\Gamma_g$ to the Riemann sphere $\CC\cup\{\infty\}$. In terms of the supersymmetric index, this is equivalent to claiming that $Z_{g,1}(T,U)^{1/24}$, seen as a function of $T$ at some fixed $U$, has only one single pole at $T\to i\infty$ and its duality images, and is holomorphic elsewhere. Since $G_g$ always contains $\Gamma_0(N)\times \Gamma_0(N)$, this amounts to claiming that $Z_{g,1}$, as a function of $T$, diverges at a given cusp of $\hat \HH/\Gamma_0(N)$ if and only if the cusp is related to $i\infty$ by an Atkin-Lehner involution in $\Gamma_g$. Next, we use the fact that $Z_{g,1}(T,U)$ is invariant (up to a sign) under T-duality along the Euclidean time direction\footnote{This T-duality corresponds to an automorphism in $O(L)$ not contained in $SO^+(L)$ and therefore it was not considered in $G_g$. Strictly speaking, the full duality duality group is generated by $G_g$ and this duality.}
\be Z_{g,1}(T,U)=-Z_{g,1}(-\frac{1}{U},-\frac{1}{T})\ .
\ee and that it is related to certain `twisted-twining' indices $Z_{g^i,g^j}$ by \be Z_{g^{a},g^{b}}(T,U)^{1/24}=Z_{g,1}(T,\frac{aU+b}{cU+d})^{1/24}=T_{1,g}(T)-T_{g^{a},g^{b}}(U)
\ .
\ee 
Using these dualities repeatedly, the behaviour of $Z_{g,1}(T,U)$ as $T$ approaches the different cusps of $\hat \HH/\Gamma_0(N)$ can be mapped to the limit of some $Z_{g^i,g^j}(T,U)$ as $U\to  \infty$ for fixed $T$. Thus, the Hauptmodul conjecture can be equivalently reformulated as the property that for  every $Z_{g^i,g^j}$ the limit $U\to \infty$ with fixed $T$ is divergent if and only if it is related to the limit $T\to \infty$, with $U$ fixed, by some duality.

\begin{figure}[h]
\includegraphics{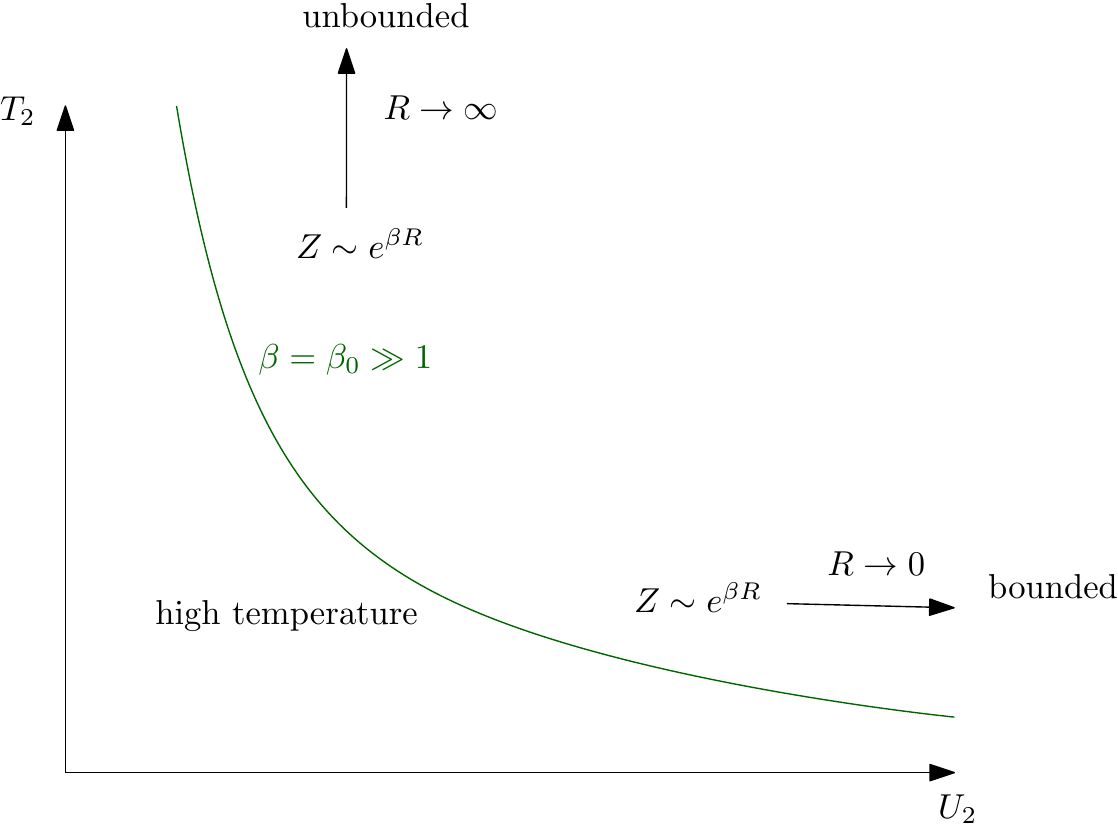}
\caption{\small The `phase diagram' for the index $Z\equiv Z_{g^i,g^j}$ at fixed $T_1=0=U_1$ in the $U_2,T_2$ plane, in the case of no phase transition. At low temperature (above the hyperbole $\beta=\beta_0\gg 1$), $Z$ is dominated by the ground state. At large radius $R$, there is always a ground state contributing $e^{\beta R}$, which diverges for $R\to \infty$ (i.e. $T_2\to \infty$, $U_2$ fixed). If there is no phase transition, then $Z\sim e^{\beta R}$ also at small radius $R\to 0$ ($T_2$ fixed, $U_2\to \infty$), so that $Z$ must be bounded in this limit.}\label{f:notrans}
\end{figure}

\begin{figure}[h]
\includegraphics{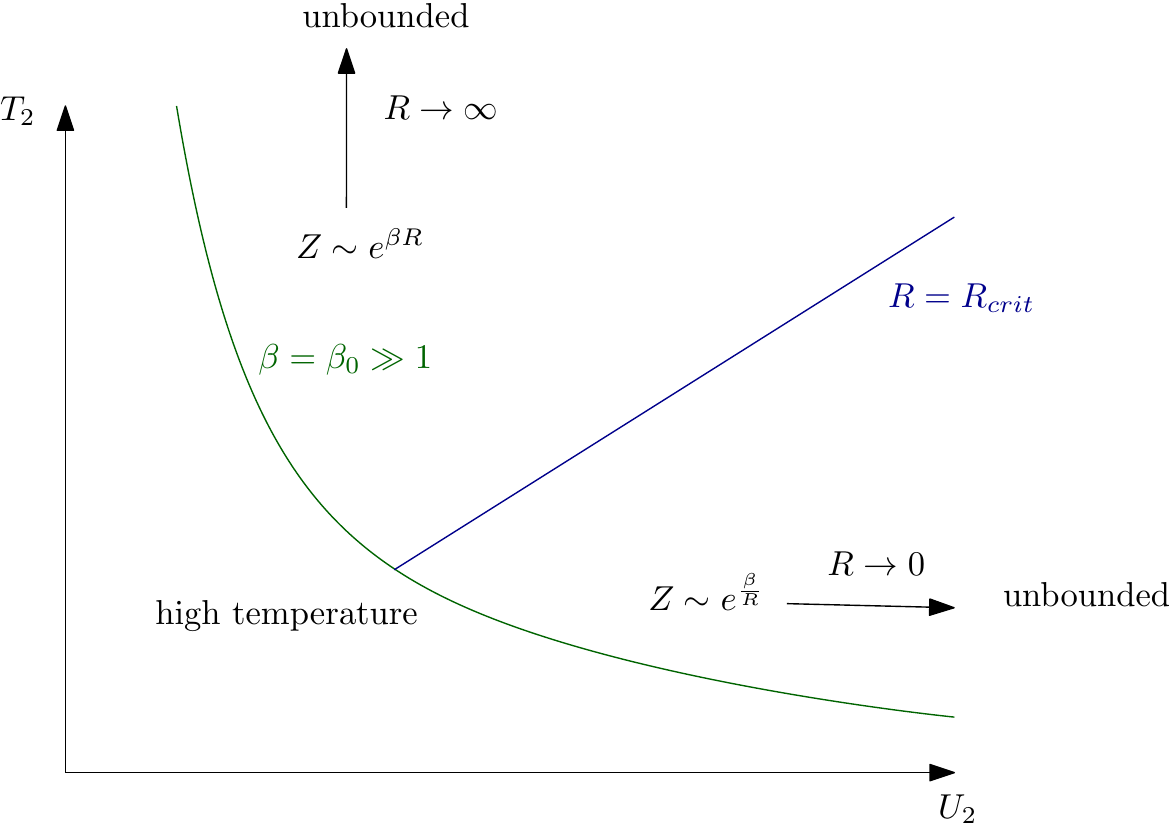}
\caption{\small The `phase diagram' for  $Z\equiv Z_{g^i,g^j}$ in the $U_2,T_2$ plane and at fixed $T_1=0=U_1$, in the case of a phase transition. At large radius, $Z$ is dominated by the ground state contribution $e^{\beta R}$. In order to have a divergence for $R\to 0$ ($T_2$ fixed, $U_2\to \infty$), a phase transition must occur: there must be a different state contributing $e^{\frac{\beta}{R}}$, which becomes dominant below a certain critical value for the radius (the line $R=R_{crit}$). On the critical line, there is an enhanced $SU(2)$ symmetry containing T-duality. Therefore, the two regions above and below the critical line are dual to each other. }\label{f:trans}
\end{figure}

It is this version of the Hauptmodul property that has a direct interpretation in our physical setup. One can smoothly interpolate between the two limits $T\to i\infty$ and $U\to i\infty$ while staying in a low temperature region $\beta\gg 1$, corresponding to $U_2,T_2\gg 1$. In this regime, the index $Z_{g^i,g^j}$ is dominated by the ground state. At large radius $R\gg 1$, there is a ground state whose contribution to $Z_{g^i,g^j}$ is $e^{\beta R}$. In the limit, $R\to \infty$, corresponding to $T\to i\infty$, this causes $Z$ to diverge. Let us now shrink the radius $R$, while keeping the temperature low. If the ground state is always the same for any value of the radius, then the $Z_{g^i,g^j}\sim e^{\beta R}$ is bounded as $R\to 0$, i.e. $U\to i\infty$ (see figure \ref{f:notrans}). So, the only way the index $Z_{g^i,g^j}$ diverges for $U\to i\infty$ is that there is some `phase transition' as the radius $R$ crosses some critical value $R_{crit}$ (see figure \ref{f:trans}): when $R<R_{crit}$ there is a different ground state whose contribution diverges as $R\to 0$ (a detailed analysis shows that this contribution must be of the form $e^{\frac{\beta}{R}}$). When the radius is exactly at the critical value, there are two degenerate ground states. In fact, one can show that the model develops an enhanced $SU(2)$ symmetry at the critical radius, under which the two degenerate states transform as a doublet. This $SU(2)$ symmetry contains in particular T-duality, exchanging deformations of the radius with different signs. This means that the two sides of the critical line are actually equivalent to each other by T-duality and the $SU(2)$ group is the usual enhanced symmetry at the self-dual radius. A more detailed description of this mechanism is given in section \ref{s:critical}.

To summarize, we showed that there are two possibilities for the index $Z_{g^i,g^j}$ in the limit $U\to i\infty$. Either the index is convergent (the case where there is no phase transition), or the index diverges and there is a self-duality relating the $U\to i\infty$ limit to the $T\to i\infty$ limit. This gives exactly the Hauptmodul property as described above.

\section{T-duality at the critical radius: a VOA interpretation}\label{s:critical}

One of the crucial steps in our physical interpretation of the Hauptmodul property is the appearance of an enhanced $SU(2)$ symmetry containing some Atkin-Lehner T-duality at some critical value of the circle in the CHL orbifold of $V^\natural\times \bar V^{s\natural}\times S^1$. This T-duality is contained in the group of self-dualities exactly when $T_{g^i,g^j}$ diverges as $U\to i\infty$. On the other hand, as explained in section \ref{sec:genus0}, an Atkin-Lehner involution $(W_e,W_e)$ arises as a T-duality if and only if the orbifold $V^\natural/\langle g^e\rangle$ is isomorphic to $V^\natural$. It is natural to wonder how these different phenomena are related to one each other.

For simplicity, we focus on the case when $T_{g,1}(U)$ diverges at $U\to i\infty$. In \cite{PPV}, using methods from string theory, we proved the following:

\begin{theorem}
Let $g\in \Aut(V^\natural)\cong M$ be an automorphism of $V^\natural$ of order $N$ and suppose that the corresponding McKay-Thompson series $T_{1,g}$ satisfies
\be\label{iltwisto} T_{g,1}(\tau)= T_{1,g}(-1/\tau)=q^{-\frac{1}{N}}+O(q^0)\ .
\ee Then, the orbifold CFT $V^\natural/\langle g\rangle$ is well-defined and is isomorphic to $V^\natural$.
\end{theorem}

Here, we reformulate the physics argument into a `sketch of a proof' in the context of vertex operator algebras. To this aim, we will use some results about orbifolds of vertex operator algebras that were recently obtained in \cite{Carnahan2016} and \cite{Ekeren:2015kq}.

Let $V_{g^n}$, $n\in \ZZ/N\ZZ$ denote the $g^n$-twisted sector (i.e. the irreducible $g^n$-twisted module) of $V^\natural$. As an ordinary reducible module for the fixed point subVOA $(V^\natural)^{\langle g\rangle}$, it can be decomposed as
\be V_{g^n}=\oplus_{m\in \ZZ/N\ZZ}  V_{n,m}\ ,
\ee where
\be V_{n,m}:=\{ v\in V_{g^n}\mid g(v)=e^{2\pi i \frac{m}{N}}\}\ ,\qquad n,m\in \ZZ/N\ZZ
\ee are the $N^2$ inequivalent irreducible modules of  $(V^\natural)^{\langle g\rangle}$. Notice that the action of $g\in \Aut(V^\natural)$ on the twisted sectors is only determined up to multiplication by a non-zero scalar. An unambiguous definition of $V_{n,m}$ requires a choice for such an action. Here, we define
\be g_{\rvert V_{g}}=e^{2\pi i L_0}\ ,
\ee on the $g$-twisted sector. On a generic $g^n$-twisted sector, $g$ is defined in such a way that the following fusion rules hold
\be V_{i,j}\boxtimes V_{k,l}\to V_{i+k,j+l}\ . 
\ee This choice is possible only when the conformal weights of the twisted sectors take values in $\frac{1}{N}\ZZ$, which is assured by the assumption \eqref{iltwisto}. More precisely, the conformal weight of $V_{n,m}$ is given by
\be\label{weights} \Delta_{V_{n,m}}=\frac{nm}{N}\mod \ZZ\ .
\ee The orbifold VOA $V^\natural/\langle g\rangle$ is given by the $g$-invariant twisted and untwisted sectors
\be V^\natural/\langle g\rangle=\oplus_{n\in \ZZ/N\ZZ} V_{n,0}\ .
\ee This is a consistent holomorphic VOA, thanks to the fact that, by \eqref{iltwisto} and \eqref{weights}, the conformal weights are all integral.

We have to prove that $V^\natural/\langle g\rangle$ is isomorphic to $V^\natural$.  To this aim, let us consider the lattice VOA $W^L$ based on the even  lattice $L=\sqrt{2N}\ZZ$ of rank $1$. Its modules $W^L_l$,  $l\in L^\vee/L\cong \ZZ/2N\ZZ$, have conformal weights
\be\label{weights2} \Delta_{W^L_l}= \frac{l^2}{4N}\mod \ZZ\ ,\qquad l\in \ZZ/2N\ZZ\ .
\ee We consider the product VOA $\tilde V_{0,0,0}:=W^L\otimes (V^\natural)^{\langle g\rangle}$ and its modules
\be \tilde V_{l,n,m} :=W^L_{l}\otimes V_{n,m}\ .
\ee  By comparing \eqref{weights} and \eqref{weights2}, it is clear that $\tilde V_{2n,n,-n}$, $n\in \ZZ/N\ZZ$, have integral conformal weights and one can consider the extended VOA
\be \oplus_{n\in \ZZ/N\ZZ} \tilde V_{2n,n,-n}\ ,
\ee and its modules
\be  \oplus_{\substack{n,m\in \ZZ/N\ZZ\\n+m=l}} \tilde V_{n-m,n,m}\ ,\qquad l\in \ZZ/N\ZZ\ .
\ee The physical idea behind this construction is the following. The CHL orbifold is defined first by tensoring $V^\natural$ with a free boson CFT on a circle $S^1$ of radius $R$, and then taking an orbifold by $(\delta,g)$, where $\delta$ is a shift by $1/N$ a period along $S^1$. The resulting theory consists of states carrying momentum $m$ and fractional winding $\frac{n}{N}$ in the free boson CFT, tensored with states in $V_{n,m}$, for all $n,m\in \ZZ$. At the radius $R=\sqrt{N}$,\footnote{In our normalization, the left- and right-moving momenta are $p_L=\frac{1}{\sqrt{2}}(\frac{m}{R}-\frac{nR}{N})$ and $p_R=\frac{1}{\sqrt{2}}(\frac{m}{R}+\frac{nR}{N})$.} the states with $m=-n$ and no right-moving oscillators are purely holomorphic ($p_R=0$) and form an extended chiral algebra, which is exactly the VOA $\oplus_{n\in \ZZ/N\ZZ} \tilde V_{2n,n,-n}$.

By \eqref{iltwisto}, the lowest weight vectors in the modules $V_{1,-1}$ and $V_{-1,1}$ have conformal weight $1-\frac{1}{N}$, while the lowest vectors in $W^L_{2}$ and $W^L_{-2}$ have weight $\frac{1}{N}$. Therefore, the vectors of lowest conformal weight of $\tilde V_{2,1,-1}$ and $\tilde V_{-2,-1,1}$ are currents $J^\pm$. Together with the current $H$ in $\tilde V_{0,0,0}\cong W^L\times (V^\natural)^{\langle g\rangle} $, they generate a $\hat{su}(2)$ Kac-Moody algebra. The module $\oplus_{\substack{n,m\in \ZZ/N\ZZ\\n+m=l}} \tilde V_{n-m,n,m}$ is therefore a module also for this $\hat{su}(2)$ algebra, with each $\tilde V_{n-m,n,m}$ being an eigenspace of $H_0$ with eigenvalue $n-m$. The $SU(2)$ group generated by the currents' zero modes contains an element $f$ whose adjoint action on the currents is
\be H\mapsto -H\qquad J^\pm\mapsto J^\mp \ .\ee
In the physical CHL orbifold description, $f$ is simply T-duality along $S^1$; this duality exists for any value of $R$, and it becomes part of the larger $SU(2)$ group at the self-dual radius $R=\sqrt{N}$.
 The element $f$ acts within each $\hat{su}(2)$-module $\oplus_{\substack{n,m\in \ZZ/N\ZZ\\n+m=l}} \tilde V_{n-m,n,m}$ and maps each $H_0$-eigenspace with eigenvalue $n-m$ to the $H_0$-eigenspace with eigenvalue $m-n$. Therefore, $f$ establishes an isomorphism of $V^L\times (V^\natural)^{\langle g\rangle}$ modules 
\be f:\tilde V_{n-m,n,m} \stackrel{\cong}{\longrightarrow} \tilde V_{m-n,m,n}\ , 
\ee compatible with fusion. In turn, this implies an isomorphism of $(V^\natural)^{\langle g\rangle}$-modules\footnote{$f$ is really of the form $f_1\otimes f_2$, where $f_1$ is the standard automorphism of the lattice VOA $W^L$ mapping $W^L_{l}$ to $W^L_{-l}$ and $f_2$ maps $V_{n,m}$ to $V_{m,n}$.}
\be f_2:V_{n,m} \stackrel{\cong}{\longrightarrow} V_{m,n}\ ,
\ee compatible with fusion. By applying the map $f_2$ to $V^\natural =\oplus_{m\in \ZZ/N\ZZ} V_{0,m}$ we obtain an isomorphism with $V^\natural/\langle g\rangle =\oplus_{n\in \ZZ/N\ZZ} V_{n,0}$.

\section{Outlook}

\noindent In this paper we give a summary of our new spacetime approach to Monstrous moonshine, first presented in \cite{PPV}. In addition to providing a gentle introduction to the more extensive work \cite{PPV}, we present a more detailed physics argument for why our model implies the genus zero property of moonshine, an explicit description of the action of the GKM algebras $\mathfrak{m}_g$ on the space of BPS states, as well as a sketch of a mathematical proof of one of the main steps of our construction using vertex operators.

An outstanding question is to better understand how our construction is related to the original BPS-algebra of Harvey and Moore \cite{Harvey:1995fq,Harvey:1996gc}. On the face of it, a central difference appears to be that we realize the space of BPS-states $\mathcal{H}_{BPS}$ as a {\it module} for the Monster Lie algebra $\mathfrak{m}$, while Harvey and Moore showed that there is an algebraic structure on $\mathcal{H}_{BPS}$, itself. However, in \cite{Harvey:1996gc} they compare the BPS-algebras of type IIA on K3 with heterotic on $T^4$ on a certain subspace of the Narain moduli space. On the type IIA side, BPS states are realized as differential forms in $H^*(\mathcal{M}(\gamma))$, where $\mathcal{M}(\gamma)$ is the moduli space of (semi-)stable sheaves on K3 in cohomology class $H^*(K3, \mathbb{Z})$. For two BPS-states $\mathcal{B}_1, \mathcal{B}_2\in H^*(\mathcal{M}(\gamma))$, the product $\mathcal{B}_1\otimes \mathcal{B}_2\rightarrow \mathcal{B}_3$ is defined using the so called \emph{correspondence conjecture} (see \cite{Harvey:1996gc}). The type IIA BPS-algebra is then identified with the Nakajima Heisenberg algebra

\begin{equation}
[\alpha_n^{I}, \alpha_m^J]=n(-1)^n \eta^{IJ}\delta_{m+n,0},
\end{equation} 

where the generators $\alpha_n^{I}$ are certain operators on $H^*(\mathcal{M}(\gamma))$ defined by Nakajima, and $\eta^{IJ}$ is the intersection pairing on $H^{*}(K3, \mathbb{Z})$. The subscript $n$  is related to the charge of the associated BPS-state $\mathcal{B}$. On the heterotic side, the operators $\alpha_n^J$ are instead realized as certain vertex operators acting on $\mathcal{H}_{BPS}$, and shown to realize a similar Heisenberg algebra. The main point here is that in this particular example, Harvey and Moore find an explicit realization of the BPS-algebra for which the space of BPS-states is a module. This is very similar in spirit to our construction.\footnote{We thank Greg Moore for discussions on this.} Given these remarks it would be very interesting to investigate whether there is a type IIA dual version of our Monstrous CHL-model. This would yield a novel connection between Monstrous moonshine and the geometry of K3-surfaces, that could possibly also shed light on the elusive relation between umbral moonshine and K3 \cite{Eguchi:2010ej,Cheng:2010pq,Gaberdiel:2010ch,Gaberdiel:2010ca,Taormina:2010pf,Cheng:2012tq,Cheng:2013wca,Creutzig:2013mqa,Cheng:2014zpa}.

\section*{Acknowledgements}
We are grateful to Richard Borcherds, Scott Carnahan, Terry Gannon, Howard Garland, and Greg Moore for discussions and correspondence. We would also like to thank the anonymous referee of \cite{PPV} for encouraging us to clarify the second quantized representation theory. We thank the organizers of the Durham Symposium on ``New Moonshine, Mock Modular Forms and String Theory'', where this work was presented. We also thank the Simons Center for Geometry and Physics for hospitality while this work was being finalized. 
NMP is supported by a National Science Foundation Graduate Research Fellowship and thanks QMAP at UC Davis for hospitality while this note was in preparation. RV is supported by a grant from Programma per Giovani Ricercatori `Rita Levi Montalcini', and thanks SLAC and the Stanford Institute for Theoretical Physics for hospitality.
\bibliographystyle{utphys}

\bibliography{Refs}

\providecommand{\href}[2]{#2}\begingroup\raggedright\begin{thebibliography}{10}

\bibitem{PPV}
N.~M. Paquette, D.~Persson, and R.~Volpato, ``Monstrous bps-algebras and the
  superstring origin of moonshine,'' {\em Communications in Number Theory and
  Physics} {\bfseries 10} no.~3, (2016) 433--526,
  \href{http://arxiv.org/abs/arXiv:1601.05412}{{\ttfamily arXiv:1601.05412}}.

\bibitem{Conway:1979kx}
J.~H. Conway and S.~P. Norton, ``Monstrous moonshine,''
  \href{http://dx.doi.org/10.1112/blms/11.3.308}{{\em Bull. London Math. Soc.}
  {\bfseries 11} no.~3, (1979) 308--339}.
  \url{http://dx.doi.org/10.1112/blms/11.3.308}.

\bibitem{Borcherds}
R.~E. Borcherds, ``Monstrous moonshine and monstrous lie superalgebras,'' {\em
  Inventiones mathematicae} {\bfseries 109} no.~1, (1992) 405--444.

\bibitem{FLM}
I.~Frenkel, J.~Lepowsky, and A.~Meurman, {\em Vertex operator algebras and the
  Monster}, vol.~134.
\newblock Academic press, 1989.

\bibitem{Norton1987}
S.~P. Norton, ``Generalized moonshine,'' in {\em The {A}rcata {C}onference on
  {R}epresentations of {F}inite {G}roups ({A}rcata, {C}alif., 1986)}, vol.~47
  of {\em Proc. Sympos. Pure Math.}, pp.~181--210.
\newblock Amer. Math. Soc., Providence, RI, 1987.
\newblock Appendix of an article by G. Mason.

\bibitem{duncan2007super}
J.~F. Duncan, ``Super-moonshine for {Conway}'s largest sporadic group,'' {\em
  Duke Mathematical Journal} {\bfseries 139} no.~2, (2007) 255--315.

\bibitem{DGH}
L.~Dixon, P.~Ginsparg, and J.~Harvey, ``Beauty and the beast: Superconformal
  symmetry in a monster module,'' {\em Communications in mathematical physics}
  {\bfseries 119} no.~2, (1988) 221--241.

\bibitem{BergmanDistler}
A.~Bergman, J.~Distler, and U.~Varadarajan, ``{(1+1) dimensional critical
  string theory and holography},''
\href{http://arxiv.org/abs/hep-th/0312115}{{\ttfamily arXiv:hep-th/0312115
  [hep-th]}}.

\bibitem{Chaudhuri:1995fk}
S.~Chaudhuri, G.~Hockney, and J.~D. Lykken, ``{Maximally supersymmetric string
  theories in $D < 10$},''
  \href{http://dx.doi.org/10.1103/PhysRevLett.75.2264}{{\em Phys.Rev.Lett.}
  {\bfseries 75} (1995) 2264--2267},
\href{http://arxiv.org/abs/hep-th/9505054}{{\ttfamily arXiv:hep-th/9505054
  [hep-th]}}.

\bibitem{Chaudhuri:1995ee}
S.~Chaudhuri and D.~A. Lowe, ``{Monstrous string-string duality},''
  \href{http://dx.doi.org/10.1016/0550-3213(96)00131-9}{{\em Nucl.Phys.}
  {\bfseries B469} (1996) 21--36},
\href{http://arxiv.org/abs/hep-th/9512226}{{\ttfamily arXiv:hep-th/9512226
  [hep-th]}}.

\bibitem{Chaudhuri:1995bf}
S.~Chaudhuri and J.~Polchinski, ``{Moduli space of CHL strings},''
  \href{http://dx.doi.org/10.1103/PhysRevD.52.7168}{{\em Phys.Rev.} {\bfseries
  D52} (1995) 7168--7173},
\href{http://arxiv.org/abs/hep-th/9506048}{{\ttfamily arXiv:hep-th/9506048
  [hep-th]}}.

\bibitem{Chaudhuri:1995dj}
S.~Chaudhuri and D.~A. Lowe, ``{Type IIA heterotic duals with maximal
  supersymmetry},'' \href{http://dx.doi.org/10.1016/0550-3213(95)00589-7}{{\em
  Nucl.Phys.} {\bfseries B459} (1996) 113--124},
\href{http://arxiv.org/abs/hep-th/9508144}{{\ttfamily arXiv:hep-th/9508144
  [hep-th]}}.

\bibitem{Persson:2015jka}
D.~Persson and R.~Volpato, ``{Fricke S-duality in CHL models},''
  \href{http://dx.doi.org/10.1007/JHEP12(2015)156}{{\em JHEP} {\bfseries 12}
  (2015) 156},
\href{http://arxiv.org/abs/1504.07260}{{\ttfamily arXiv:1504.07260 [hep-th]}}.

\bibitem{2008arXiv0812.3440C}
S.~{Carnahan}, ``{Generalized Moonshine I: Genus zero functions},'' {\em ArXiv
  e-prints} (Dec., 2008) , \href{http://arxiv.org/abs/0812.3440}{{\ttfamily
  arXiv:0812.3440 [math.RT]}}.

\bibitem{Carnahan2014}
S.~Carnahan, ``Generalized moonshine, {II}: {B}orcherds products,''
  \href{http://dx.doi.org/10.1215/00127094-1548416}{{\em Duke Math. J.}
  {\bfseries 161} no.~5, (2012) 893--950},
  \href{http://arxiv.org/abs/0908.4223}{{\ttfamily arXiv:0908.4223}}.
  \url{http://dx.doi.org/10.1215/00127094-1548416}.

\bibitem{2012arXiv1208.6254C}
S.~{Carnahan}, ``{Generalized Moonshine IV: Monstrous Lie algebras},'' {\em
  ArXiv e-prints} (Aug., 2012) ,
  \href{http://arxiv.org/abs/1208.6254}{{\ttfamily arXiv:1208.6254 [math.RT]}}.

\bibitem{Harvey:1995fq}
J.~A. Harvey and G.~W. Moore, ``{Algebras, BPS states, and strings},''
  \href{http://dx.doi.org/10.1016/0550-3213(95)00605-2}{{\em Nucl.Phys.}
  {\bfseries B463} (1996) 315--368},
\href{http://arxiv.org/abs/hep-th/9510182}{{\ttfamily arXiv:hep-th/9510182
  [hep-th]}}.

\bibitem{Harvey:1996gc}
J.~A. Harvey and G.~W. Moore, ``{On the algebras of BPS states},''
  \href{http://dx.doi.org/10.1007/s002200050461}{{\em Commun.Math.Phys.}
  {\bfseries 197} (1998) 489--519},
\href{http://arxiv.org/abs/hep-th/9609017}{{\ttfamily arXiv:hep-th/9609017
  [hep-th]}}.

\bibitem{Neumann:1997pr}
C.~D.~D. Neumann, ``{Perturbative BPS algebras in superstring theory},''
  \href{http://dx.doi.org/10.1016/S0550-3213(97)00360-X}{{\em Nucl. Phys.}
  {\bfseries B499} (1997) 596--620},
\href{http://arxiv.org/abs/hep-th/9702197}{{\ttfamily arXiv:hep-th/9702197
  [hep-th]}}.

\bibitem{Fiol:2000wx}
B.~Fiol and M.~Marino, ``{BPS states and algebras from quivers},''
  \href{http://dx.doi.org/10.1088/1126-6708/2000/07/031}{{\em JHEP} {\bfseries
  07} (2000) 031},
\href{http://arxiv.org/abs/hep-th/0006189}{{\ttfamily arXiv:hep-th/0006189
  [hep-th]}}.

\bibitem{Kontsevich:2010px}
M.~Kontsevich and Y.~Soibelman, ``{Cohomological Hall algebra, exponential
  Hodge structures and motivic Donaldson-Thomas invariants},''
  \href{http://dx.doi.org/10.4310/CNTP.2011.v5.n2.a1}{{\em Commun. Num. Theor.
  Phys.} {\bfseries 5} (2011) 231--352},
\href{http://arxiv.org/abs/1006.2706}{{\ttfamily arXiv:1006.2706 [math.AG]}}.

\bibitem{Gukov:2011ry}
S.~Gukov and M.~Sto{\v s}i{\'c}, ``{Homological Algebra of Knots and BPS
  States},'' \href{http://dx.doi.org/10.1090/pspum/085/1377}{{\em Proc. Symp.
  Pure Math.} {\bfseries 85} (2012) 125--172},
\href{http://arxiv.org/abs/1112.0030}{{\ttfamily arXiv:1112.0030 [hep-th]}}.

\bibitem{Chuang:2013wt}
W.-y. Chuang, D.-E. Diaconescu, J.~Manschot, G.~W. Moore, and Y.~Soibelman,
  ``{Geometric engineering of (framed) BPS states},''
  \href{http://dx.doi.org/10.4310/ATMP.2014.v18.n5.a3}{{\em Adv. Theor. Math.
  Phys.} {\bfseries 18} no.~5, (2014) 1063--1231},
\href{http://arxiv.org/abs/1301.3065}{{\ttfamily arXiv:1301.3065 [hep-th]}}.

\bibitem{Gaberdiel:2011qu}
M.~R. Gaberdiel, S.~Hohenegger, and D.~Persson, ``{Borcherds Algebras and N=4
  Topological Amplitudes},''
  \href{http://dx.doi.org/10.1007/JHEP06(2011)125}{{\em JHEP} {\bfseries 06}
  (2011) 125},
\href{http://arxiv.org/abs/1102.1821}{{\ttfamily arXiv:1102.1821 [hep-th]}}.

\bibitem{Hohenegger:2011ff}
S.~Hohenegger and D.~Persson, ``{Enhanced Gauge Groups in N=4 Topological
  Amplitudes and Lorentzian Borcherds Algebras},''
  \href{http://dx.doi.org/10.1103/PhysRevD.84.106007}{{\em Phys. Rev.}
  {\bfseries D84} (2011) 106007},
\href{http://arxiv.org/abs/1107.2301}{{\ttfamily arXiv:1107.2301 [hep-th]}}.

\bibitem{kostant}
B.~Kostant, ``Lie algebra cohomology and the generalized borel-weil theorem,''
  {\em Annals of Mathematics} (1961) 329--387.

\bibitem{nuovo}
D.~Persson and R.~Volpato, ``Dualities in {CHL} models,'' {\em to appear} .

\bibitem{Carnahan2016}
S.~{Carnahan} and M.~{Miyamoto}, ``{Regularity of fixed-point vertex operator
  subalgebras},'' {\em ArXiv e-prints} (Mar., 2016) ,
  \href{http://arxiv.org/abs/1603.05645}{{\ttfamily arXiv:1603.05645
  [math.RT]}}.

\bibitem{Ekeren:2015kq}
J.~van Ekeren, S.~M{\"o}ller, and N.~R. Scheithauer, ``Construction and
  classification of holomorphic vertex operator algebras,''
  \href{http://arxiv.org/abs/1507.08142}{{\ttfamily arXiv:1507.08142
  [math.RT]}}.

\bibitem{Eguchi:2010ej}
T.~Eguchi, H.~Ooguri, and Y.~Tachikawa, ``{Notes on the K3 Surface and the
  Mathieu group $M_{24}$},''
  \href{http://dx.doi.org/10.1080/10586458.2011.544585}{{\em Exper.Math.}
  {\bfseries 20} (2011) 91--96},
\href{http://arxiv.org/abs/1004.0956}{{\ttfamily arXiv:1004.0956 [hep-th]}}.

\bibitem{Cheng:2010pq}
M.~C. Cheng, ``{K3 Surfaces, N=4 Dyons, and the Mathieu Group M24},''
  \href{http://dx.doi.org/10.4310/CNTP.2010.v4.n4.a2}{{\em
  Commun.Num.Theor.Phys.} {\bfseries 4} (2010) 623--658},
\href{http://arxiv.org/abs/1005.5415}{{\ttfamily arXiv:1005.5415 [hep-th]}}.

\bibitem{Gaberdiel:2010ch}
M.~R. Gaberdiel, S.~Hohenegger, and R.~Volpato, ``{Mathieu twining characters
  for K3},'' \href{http://dx.doi.org/10.1007/JHEP09(2010)058}{{\em JHEP}
  {\bfseries 1009} (2010) 058},
\href{http://arxiv.org/abs/1006.0221}{{\ttfamily arXiv:1006.0221 [hep-th]}}.

\bibitem{Gaberdiel:2010ca}
M.~R. Gaberdiel, S.~Hohenegger, and R.~Volpato, ``{Mathieu Moonshine in the
  elliptic genus of K3},''
  \href{http://dx.doi.org/10.1007/JHEP10(2010)062}{{\em JHEP} {\bfseries 1010}
  (2010) 062},
\href{http://arxiv.org/abs/1008.3778}{{\ttfamily arXiv:1008.3778 [hep-th]}}.

\bibitem{Taormina:2010pf}
A.~Taormina and K.~Wendland, ``{The Symmetries of the tetrahedral Kummer
  surface in the Mathieu group $M_24$},''
\href{http://arxiv.org/abs/1008.0954}{{\ttfamily arXiv:1008.0954 [hep-th]}}.

\bibitem{Cheng:2012tq}
M.~C. Cheng, J.~F. Duncan, and J.~A. Harvey, ``{Umbral Moonshine},''
  \href{http://dx.doi.org/10.4310/CNTP.2014.v8.n2.a1}{{\em
  Commun.Num.TheorPhys.} {\bfseries 08} (2014) 101--242},
\href{http://arxiv.org/abs/1204.2779}{{\ttfamily arXiv:1204.2779 [math.RT]}}.

\bibitem{Cheng:2013wca}
M.~C. Cheng, J.~F. Duncan, and J.~A. Harvey, ``Umbral moonshine and the
  niemeier lattices,'' \href{http://dx.doi.org/10.1186/2197-9847-1-3}{{\em
  Research in the Mathematical Sciences} {\bfseries 1} no.~1, (2014) 1--81},
  \href{http://arxiv.org/abs/1307.5793}{{\ttfamily arXiv:1307.5793 [math.RT]}}.
\url{http://dx.doi.org/10.1186/2197-9847-1-3}.

\bibitem{Creutzig:2013mqa}
``{Mathieu Moonshine and the Geometry of K3 Surfaces},''
  \href{http://dx.doi.org/10.4310/CNTP.2014.v8.n2.a3}{{\em
  Commun.Num.TheorPhys.} {\bfseries 08} (2014) 295--328},
\href{http://arxiv.org/abs/1309.2671}{{\ttfamily arXiv:1309.2671 [math.QA]}}.

\bibitem{Cheng:2014zpa}
M.~C.~N. Cheng and S.~Harrison, ``{Umbral Moonshine and K3 Surfaces},''
  \href{http://dx.doi.org/10.1007/s00220-015-2398-5}{{\em Commun. Math. Phys.}
  {\bfseries 339} no.~1, (2015) 221--261},
\href{http://arxiv.org/abs/1406.0619}{{\ttfamily arXiv:1406.0619 [hep-th]}}.

\end{thebibliography}\endgroup

\end{document}